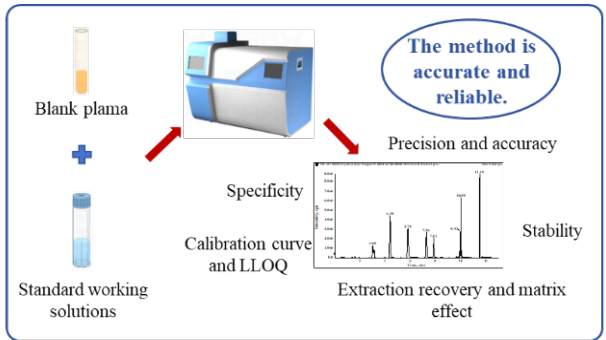
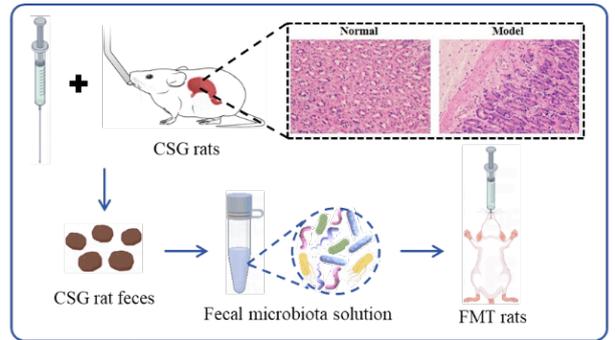
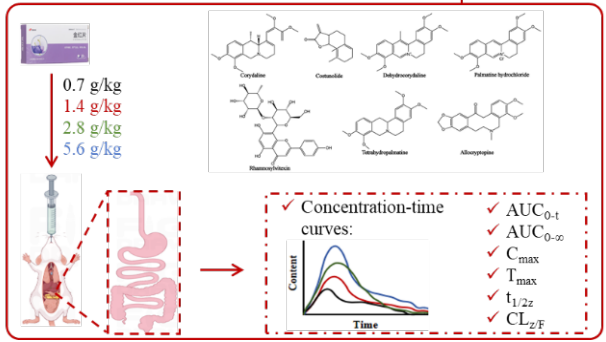
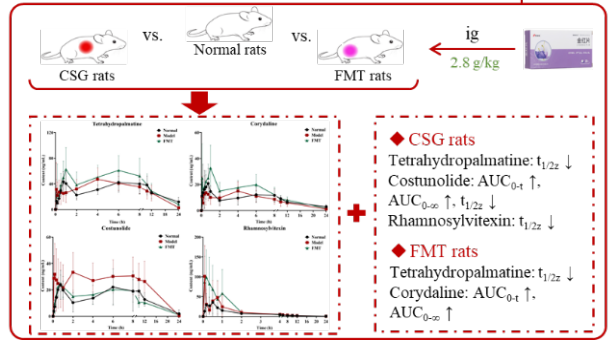

**Graphical abstract**

# Pharmacokinetic characteristics of Jinhong tablets in normal, chronic superficial gastritis and intestinal microbial disorder rats

Tingyu Zhang [a, #], Jian Feng [b, c, #], Xia Gao [b, c], Xialin Chen [b, c], Hongyu Peng [b, c], Xiaoxue Fan [b, c], Xin Meng [b, c], Mingke Yin [b, c], Zhenzhong Wang [b, c], Bo Zhang [a, *], Liang Cao [b, c, *]

[a] Institute for TCM-X, Department of Automation, Tsinghua University, Beijing, 100084, China

[b] State Key Laboratory of Technologies for Chinese Medicine Pharmaceutical Process Control and Intelligent Manufacture, Jiangsu Kanion Pharmaceutical Co.,Ltd., Jiangsu, Nanjing, 210000, China

[c] Jiangsu Kanion Modern Chinese Medicine Institute, Nanjing, 211100, China

#: These authors contributed equally.

*: Corresponding author:

Liang Cao, State Key Laboratory of Technologies for Chinese Medicine Pharmaceutical Process Control and Intelligent Manufacture, Jiangsu Kanion Pharmaceutical Co. Ltd, Lianyungang, 222001, China.

E-mail address: leancao@163.com

## Abstract

**Enthnopharmacological relevance:** Jinhong tablet (JHT), a traditional Chinese patent medicine prepared from four Chinese medicinal materials, has good efficacy of soothing the liver and relieving depression, regulating qi and promoting blood circulation, regulating stomach and relieving pain, clinically treating chronic superficial gastritis (CSG) with disharmony between liver and stomach. However, its pharmacokinetics research is limited.

**Aim of the study:** This study aims to explore the pharmacokinetic characteristics of JHT in normal rats and reveal its pharmacokinetic difference in normal, CSG and intestinal microbial disorder rats.

**Materials and methods:** A quantitative analytical method was established for seven active ingredients in rat plasma using UPLC-TQ-MS/MS. After oral administrated four different doses of JHT, the plasma concentration of the seven ingredients were determined to assess pharmacokinetic characteristics in normal rats. Subsequently, Pharmacokinetic difference of four main ingredients was observed in normal, CSG, and fecal microbiota transplantation (FMT) rats after JHT-administrated. Simultaneously, the intestinal microbial changes of the three groups were evaluated by high-throughput sequencing technique. Finally, Spearman correlation analysis was applied to characterize the connection between the in *vivo* exposure of four ingredients and the disturbed gut microbiota.

**Results:** A appropriate quantitative method was successfully established with the good linearity, precision, accuracy, extraction recovery, matrix effect as well as stability. In normal rats, all seven ingredients were rapidly absorbed. Tetrahydropalmatine, corydaline, costunolide, and rhamnosylvitexin demonstrated good exposure, while dehydrocorydaline, allocryptopine, and palmatine hydrochloride had low exposure. Additionally, tetrahydropalmatine, corydaline, and costunolide exhibited linear pharmacokinetics in terms of $AUC_{0-t}$ and $C_{max}$ when administrated 0.7~5.6 g/kg of JHT, whereas rhamnosylvitexin and dehydrocorydaline showed linear pharmacokinetics within 0.7~2.8 g/kg. In CSG and FMT rats, the pharmacokinetic behavior of JHT displayed differences. CSG significantly enhance the *in vivo* exposure and $C_{max}$ of costunolide, with increasing those of rhamnosylvitexin. FMT markedly raised exposure of corydaline and $C_{max}$ of rhamnosylvitexin, closely related to 20 differential bacterial genera.

**Conclusions:** Jinhong tablets exhibited the linear pharmacokinetic characteristics within the appropriate dose ranges and the CSG pathological state and intestinal microbial disorder obviously affect its absorption and metabolism. These findings provide the meaningful information for the mechanism research and clinical application of JHT.

# Abbreviations

JHP: jinhong tablet

FMT: fecal microbiota transplantation

CSG: chronic superficial gastritis

AUC: area under the plasma concentration-time curve

$CL_{z/F}$: clearance

$T_{max}$: time to reach the maximum concentration

$C_{max}$: maximum plasma concentration

$T_{1/2z}$: half-life

DP: declustering potential

CE: collision energy

ESI: electrospray ionization

IS: internal standard

LLOQ: lower limit of quantification

MRM: multiple reaction monitoring

QC: quality control

RE: relative error

RSD: relative standard deviation

SD: standard deviation

TCM: traditional Chinese medicine

SD rats: Sprague-Dawley rats

P-gp: P-glycoprotein

PLS-DA: partial least squares discriminant analysis

Q1: precursor ion

Q3: product ion

UPLC-TQ-MS/MS: ultra-high performance liquid chromatography coupled with triple quadrupole mass spectrometry

# 1. Introduction

Chronic superficial gastritis (CSG), also known as chronic non-atrophic inflammation, is a common chronic disease in the digestive system, which mainly manifests as inflammation of the superficial layer of gastric mucosa. Most of patients have no specific symptoms, and those with symptoms mainly appear as persistent or recurrent epigastric pain, abdominal distension and satiety, similar to those with dyspepsia [Zhao et al, 2024]. There is multiple pathogenesis, including Helicobacter pylori (H. pylori) infection, alcohol consumption, bile reflux, and autoimmune gastritis [Xu et al, 2022]. With the socioeconomic development and the people's dietary structure change, the incidence of CSG occupies the first place in chronic stomach diseases, and 80% to 90% of patients undergoing gastroscopy are diagnosed as CSG [Fang and Zou, 2021].

Network pharmacology is an emerging interdisciplinary field that integrates systems biology, bioinformatics, and pharmacology to elucidate the complex interactions between drugs, targets, and diseases from a network perspective [Lai et al., 2020; Su et al., 2012]. Unlike traditional reductionist approaches, network pharmacology provides a holistic framework to analyze multi-component, multi-target therapeutic systems, making it particularly suitable for studying traditional Chinese medicine (TCM) [张 and 李, 2015; 李, 2011]. By constructing compound-target-disease networks, this approach enables the systematic identification of key bioactive compounds within TCM formulations that drive pharmacological effects [B. Wang et al., 2024; Z. Wang et al., 2024; Zhang et al., 2023]. Such a strategy not only accelerates the discovery of active ingredients but also bridges the gap between computational predictions and experimental pharmacology [Li et al., 2006, 2003].

Jinhong tablet (JHT) is an innovative Chinese patent medicine, which originated from Jin-Ling-Zi San in "Su Wen Bing Ji Qi Yi Bao Ming Ji" written by Wansu Liu, a famous ancient Chinese physician. JHT is manufactured by four herbal medicines in a fixed mass ratio of 1.5:1.5:1:1, including *Melia azedarach* L. (Chuanlianzi), *Corydalis yanhusuo* (Y.H.Chou & Chun C.Hsu) W.T.Wang ex Z.Y.Su & C.Y.Wu (Yanhusuo), *Dolomiaea souliei* (Franch.) C.Shih (Chuanmuxiang), and *Illicium dunnianum* Tutch. (Honghuabajiaoye). All the plant names have been checked with MPNS (https://mpns.science.kew.org/). In the prescription, chuanlianzi is used for clearing liver fire, with yanhusuo for regulating qi, chuanmuxiang for promoting qi and relieving pain, and honghuabajiaoye for dispersing blood stasis and promoting qi. For the efficacy of soothing the liver and relieving depression, regulating qi and promoting blood circulation, regulating stomach and relieving pain, JHT is clinically used to treating CSG with disharmony between liver and stomach (symptoms include epigastric pain, acid swallowing, belching, hiccups, etc.). Several clinical trials proved that JHT had definite and remarkable curative effect in CSG therapy [Chen, 2010; Wei et al, 2000; Zhang et al, 1998]. In addition, pharmacological studies illustrated that JHT could reduce the total acid output, inhibit pepsin secretion, retard gastric emptying, and promote ulcer healing, and resist gastric ulcer, with analgesic, spasmolytic, anti-inflammatory, and anti-gastric ulcer effects [Zhang et al, 1998; Liu et al, 2023].

In our previous study, 96 chemical constituents were identified in JHT, and 163 xenobiotics (38 prototypes and 125 metabolites) were characterized in rat after oral administration of JHT using ultra-performance liquid chromatography coupled with quadrupole time-of-flight tandem mass spectrometry, mainly including alkaloids, organic acids, flavonoids, and others [Liu et al, 2021a]. Based on these, 12 components (tetrahydropalmatine, corydaline, palmatine, allocryptopine, protopine, shikimic acid, chlorogenic acid, isochlorogenic acid B, isochlorogenic acid C, quercetin, isotoosendanin, and toosendanin) were simultaneously determined using ultra-performance liquid chromatography coupled with a triple-stage quadrupole mass spectrometer (UPLC-TQ-MS/MS) [Liu et al, 2021b]. Among them, tetrahydropalmatine and corydaline were applied for analgesia [Guan et al, 2020] and adjuvant therapy of digestive tract ulcer [Zeng et al, 2017]; costunolide and dehydrocostunolide had pharmacological activities of promoting gastric emptying and inhibiting digestive tract ulcer [Wang et al, 2020a; Mao et al, 2017]; rhamnosylvitexin not only alleviated chronic inflammation by regulating intestinal flora [Huang et al, 2024], but also contributed to the protection against oxidative stress damage [Wei et al, 2014]. However, its pharmacokinetics has not been studied.

For complex traditional Chinese medicine, the confirmation of vital chemical constituents, with biological activity, safety, pharmacokinetic properties, and high content, is conducive to its quality control and whole pharmaceutical process traceability, better reflecting the essence of TCM in treating diseases [Li et al, 2021a]. Pharmacokinetic research can screen out the potential active components of TCM, reveal their dynamic process *in vivo*, and thus clarify the vital constituents related to curative effect [Yan et al, 2018; Hu et al, 2013]. Furthermore, it is well known that gastric pathological changes and intestinal microbial disorder easily lead to impaired gastric acid secretion [Fiorini et al, 2015; Lahner et al, 2009], destroyed gastrointestinal mucosal barrier [Chen et al, 2025] and changed metabolic enzyme activity [Huang et al, 2023; Tsunoda et al, 2021], causing unintended variations in drug absorption and metabolism. Moreover, chronic gastritis can also induce intestinal microbial disorder [Bu et al, 2024; Chen et al, 2024]. Therefore, it is necessary to compare the pharmacokinetic characteristics of JHT under CSG and intestinal microbial disorder.

Based on our preliminary experiments, a highly sensitive and accurate quantitative method for the seven bioactive components in rat plasma was developed using UPLC-TQ-MS/MS. Then, the pharmacokinetics features of JHT in rats were evaluated by gavage with four different doses. Subsequently, the exposure levels of its key bioactive components were compared in three different physiological states, including normal, CSG and intestinal microbial disorder rats. Notably, elucidating the pharmacokinetics differences of JHT under different doses and physiological conditions play a crucial significance for rational clinical use and the pharmacokinetic-pharmacodynamic correlation research.

## 2. Materials and methods

### 2.1 Experimental reagents

JHTs were supplied by Jiangsu Kangyuan Pharmaceutical Co., Ltd (batch No. 230602, the contents of tetrahydropalmatine, corydaline, dehydrocorydaline, costunolide, rhamnosylvitexin, palmatine hydrochloride, and allocryptopine were 0.77, 0.98, 1.29, 0.14, 1.82, 0.18, and 0.30 mg/g, respectively; their structures were as shown in Fig. 1). Allocryptopine (batch No. J21HB174666, purity 98%), corydaline (batch No. M25HB179211, purity 98%), and dehydrocorydaline (batch No. A16HB191857, purity 97.5%) were purchased from Shanghai Yuan Ye Biotechnology Co. Ltd. Palmatine hydrochloride (batch No. 110732-201913, purity 85.7%), tetrahydropalmatine (batch No. 110726-202020, purity 99.3%), rhamnosylvitexin (batch No. 111668-200602, purity 100%), costunolide (batch No. 111524-202312, purity 99.6%) and hydrocortisone (batch No. 100152-202008, purity 99.2%) were purchased from China Academy of Food and Drug Testing and Research. Formic acid (Thermo Fisher Scientific (China) Co., Ltd, batch No. 220622). Methanol and acetonitrile were pure for mass spectrometry, and water was purified water.

### 2.2 Animal experiments

#### 2.2.1 Experimental animals and ethics statement

Male Sprague-Dawley (SD) rats (weighing 200-220 g) were provided by Hangzhou Medical College and Beijing Vital River Laboratory Animal Technology Co., Ltd., with production licenses SCXK (Zhe) 2019-0002 and SCXK (Jing) 2021-0006, respectively. All rats were housed in an environment with a temperature of 23°C±2°C, relative humidity of 60%±10%, and a 12-hour light/12-hour dark cycle. This study was strictly conducted in accordance with the recommendations in the Guide for the Care and Use of Laboratory Animals of the National of Institutes Health (NIH Publication No. 85-23, revised in 1985). The breeding and experimental operations of the animals strictly followed the regulations of the Laboratory Animal Ethics Committee of Jiangsu Kanion Pharmaceutical Co., Ltd., with the animal use license number SYXK (Su) 2023-0035. All animal experimental protocols were approved by the Institutional Animal Care and Use Committee (IACUC) of Jiangsu Kanion Pharmaceutical Co., Ltd (Approval NO. 2024022905).

2.2.2 Pharmacokinetic studies of JHT in normal rats after administrated four different doses

Thirty-two male rats were randomly divided into four groups, with eight rats in each group. The rats were fasted for 12 h before JHT-administration and were given free water. Each group of rats were administered JHT at the dosage of 0.7, 1.4, 2.8, and 5.6 g/kg (1, 2, 4, and 8 times equivalent to the oral dose of humans, respectively). Blood samples were collected from the orbital vein of the rats at 0 min (before administration), 5 min, 10 min, 20 min, 30 min, 45 min, 1 h, 2 h, 4 h, 6 h, 8 h, 10 h, 12 h, and 24 h after JHTs were administrated. Next, plasma was obtained by centrifugation at 8,000 rpm for 5 min and stored at -80ºC.

2.2.3 Fecal microbiota solution preparation

Fresh feces were collected from rats in the model group, and fecal microbiota suspension was prepared within 2 hours after defecation. First, the feces were added to PBS at the ratio of 1:10 (w:v) and stirred well by homogenizer. Next, samples passed through a 200-mesh sterilizing grade filter, followed by centrifugation at 3000 r/min for 15 min to collect the fecal bacteria extract. Finally, all extracts were added an equal proportion of sterile PBS for gavage to rats.

2.2.4 Establishment of CSG rats and FMT rats

Twenty-six healthy male rats were randomly divided into three groups, eight in normal group, ten in model group, and eight in FMT group. Rats form the model group were subjected to alternating intragastric gavage with a mixture of 30% ethanol and 2% sodium salicylate (three times weekly, 10 mL/kg) and 10 mM sodium deoxycholate solution (twice weekly, 10 mL/kg). Additionally, rats in the model group were provided with 0.05% aqueous ammonia as drinking water and subjected to an irregular diet (two-day full feeding, one-day fasting). The construction of the CSG model lasted 57 days. On the 31th day of modeling, rats in the FMT group were provided with water mixed with antibiotics (containing 0.5 g/L vancomycin hydrochloride, 1 g/L penicillin sodium, 1 g/L streptomycin sulfate, 1 g/L neomycin sulfate, and 1 g/L metronidazole) for 17 days. The administration of fecal bacteria solution was then initiated by gavage at 1 mL/d until the end of the model. After the modeling ended, the feces of rats in each group were collected and stored at -80°C.

2.2.5 Plasma and tissue collection from normal, CSG, and FMT rats after JHT-administration

All rats of the normal, CSG, and FMT groups were administered a single dose of JHT at 2.8 g/kg. Blood samples were collected at 0 min (before administration), 5 min, 10 min, 20 min, 30 min, 45 min, 1 h, 2 h, 4 h, 6 h, 8 h, 10 h, 12 h, and 24 h after JHT-administration. Plasma was obtained by centrifugation at 8,000 rpm for 5 min and stored at −80ºC. After plasma collection, the rats were euthanized using a 10% urethane solution. Subsequently, the abdominal cavity of the rats was opened, and the gastric tissues were separated. Then gastric contents were removed, and the gastric tissues were washed with physiological saline and then immersed in 4% paraformaldehyde solution for fixation.

Scoring criteria: 0: No inflammatory cell infiltration; 1: Multiple chronic inflammatory cells are visible in the bottom of the gastric mucosal epithelium or lamina propria; 2: Numerous inflammatory cells infiltrate from the gastric mucosal epithelial layer to the muscularis mucosae; 3: Clusters of inflammatory cells are visible within the gastric mucosa. Where between the above levels, the number of the lower level was added 0.5, and 3 fields of view were selected from each slice for observation and scoring to evaluate the infiltration of inflammatory cells.

2.3 Plasma sample preparation

100 µL of plasma was mixed with 400 µL of ice-cold internal standard solution. The samples were vortexed for 1 min, and centrifuged at 12000 rpm for 10 min. The supernatant was transferred to a new tube and evaporated to dryness at 37 ºC under a gentle stream of nitrogen. 100 µL of 20% methanol solution to re-dissolve the residue, vortex for 1 min, and centrifuge at 12000 rpm for 10 min. Then take the supernatant into the sample for analysis.

2.4 UPLC-TQ-MS/MS conditions for plasma sample detection

A Shimadzu LC-40AD UPLC (Shimadzu Corporation, Kyoto, Japan) and an AB API6500+ triple-quadrupole mass spectrometer (AB Sciex, Framingham, Massachusetts, USA) equipped with an ESI interface were used in this study. Multiple reaction monitoring (MRM) mode was used to quantify the seven ingredients. Seven ingredients were detected in positive ion mode (ESI$^+$). The operating parameters were set as follows: ion spray voltage: 5500 V; source temperature: 500 ºC; ion source gas I, gas II, collision gas, and curtain gas: 50, 50, 9, and 35 psi, respectively. The precursor/product ion pairs (Q1/Q3), declustering potential (DP), and collision energy (CE) of seven ingredients are listed in Table S1. The acquisition and analysis of data were performed using AB SCIEX LC/MS Analyst software, version 1.6.3.

Seven ingredients were separated using Agilent Eclipse Plus $C_{18}$ column (2.1 mm × 100 mm, 1.8 µm; Agilent, Palo Alto, California, USA) under the column temperature of 35 ºC. The mobile phase consisted of 0.1% formic acid (A) and acetonitrile (B) at a flow rate of 0.4 mL/min. The gradient elution was set as follows: 0~4.5 min, 22% B; 4.5~8.5 min, 22%~30% B; 8.5~8.7 min, 30%~60% B; 8.7~10.5 min, 60% B; 10.5~10.7 min, 60%~90% B; 10.7~11.5 min, 90% B; 11.5~12 min, 90%~22% B; 12~13 min, 22% B. The injection volume was 3 µL.

2.5 Preparation of standard working solutions and internal standards

Tetrahydropalmatine, corydaline, dehydrocorydaline, costunolide, rhamnosylvitexin, and allocryptopine were weighed accurately, and dissolved in methanol to prepare the stock solutions of 1 mg/mL. Accurately weigh palmatine hydrochloride and prepare a stock solution with a concentration of 1 mg/mL using 50% methanol. Mix an appropriate amount of ingredient stock solutions and dilute them with methanol in a 60% gradient to obtain a series of standard working solutions. The standard working solutions were spiked into blank plasma, and processed as described in "2.3 Plasma sample preparation" to obtain calibration samples. The concentration ranges for seven ingredients were as follows: tetrahydropalmatine and corydaline, 40~1.2 ng/mL; palmatine hydrochloride, 100~2.8 ng/mL; costunolide, 250~7.0 ng/mL; allocryptopine and rhamnosylvitexin, 25~0.7 ng/mL; and dehydrocorydaline, 20~0.56 ng/mL.

Hydrocortisone (IS) was weighed accurately, and dissolved in methanol to prepare the stock solutions of 1 mg/mL. Appropriate amount of hydrocortisone stock solution was aspirated, mixed with an appropriate amount of mass spectrometry-grade formic acid, and then diluted with methanol to prepare a solution with a mass concentration of 62.5 ng/mL and a formic acid content of 0.125%.

2.6 Method validation

2.6.1 Specificity

Compare the blank plasma, plasma samples spiked with the lower limit of quantitation (LLOQ) mixture, and plasma samples after oral administration of JHT to assess the specificity of the method.

2.6.2 Calibration curve and LLOQ

In accordance with the plasma sample processing method, blank plasma spiked with a series of mixed standard solutions was treated. A calibration curve was then constructed, with the ratio of the peak area of the ingredient to the peak area of the internal standard (A/Ai) as the Y-coordinate, and the mass concentration of each ingredient (X) as the X-coordinate.

2.6.3 Precision and accuracy

LLOQ and low, medium, and high concentration quality control (QC) samples were analyzed to assess intra-day and inter-day precision and accuracy. Six samples in parallel for each concentration were injected consecutively during the day, and three consecutive analytical batches were measured during the day, with a minimum interval of 12 h between each analytical batch.

The accuracy (expressed as relative error, RE) and precision (expressed as relative standard deviation, RSD) of the LLOQ samples should be less than 20%. For QC samples, the RSD and RE should both be within 15%.

2.6.4 Extraction recovery and matrix effect

The extraction recovery and matrix effect of the ingredients was evaluated at low, medium, and high concentrations quality control. The extraction recovery was determined by comparing the peak areas of ingredients extracted from spiked plasma samples with those of ingredients in unextracted spiked plasma samples. Matrix effect of seven ingredients were investigated by calculating the ratio of the mean peak area of the ingredients in unextracted spiked plasma samples to the peak area of the pure standard solution at the same concentration.

2.6.5 Stability

The stability of the seven ingredients was evaluated by analyzing low and high concentrations quality control (QC) samples under various conditions, including placement in the autosampler (4°C) for 24 hours, storage at room temperature for 4 hours, long-term storage at -80°C for 30 days, and three freeze-thaw cycles (freezing at -80°C and thawing at room temperature). Six samples were parallelized for each concentration.

2.7 Statistical analysis

The pharmacokinetic parameters were estimated using the DAS 3.2.8 software. The pharmacokinetic parameters were calculated using the statistical moment method, and statistical analysis was performed. The pharmacokinetic parameters were expressed as mean ± standard deviation (SD).

**3. Result**

3.1 Method validation

The Specificity results are summarized in Fig. 1. The results show good separation between the seven ingredients and the internal standard, no endogenous interference was observed, and the interference peak response of the target ingredient in the blank plasma sample was less than 20% of the LLOQ, and the interference peak response of the internal standard was less than 5% of the LLOQ (Table S2).

The results of calibration curve and LLOQ are listed in Table S3, and the seven ingredients in rat plasma were good linear within their linear ranges ($r > 0.99$). The LLOQ was 1.12 ng/mL for tetrahydropalmatine and corydaline, 2.80 ng/mL for palmatine hydrochloride, 7.00 ng/mL for costunolide, 0.70 ng/mL for allocryptopine and rhamnosylvitexin, and 0.56 ng/mL for dehydrocorydaline.

The intra-day and inter-day precision of seven ingredients is within 15%, and the RE is between -13.62% and 8.53% (Table S4), indicating that the method is accurate, reliable and reproducible, and meets the requirements of the analytical method for biological samples.

The results of extraction recovery and matrix effect are shown in Table S5, the extraction recoveries of the seven ingredients were between 68% and 87% with RSD less than 14.75%, and the matrix effect ranged from 0.71 to 1.19 with RSD less than 14.82%, which indicates that the method has good extraction recovery and little interference from matrix effect.

The results of stability are shown in Table S6, the seven ingredients of rat plasma maintained at room temperature for 4 h (RE, -14.96%~14.44%), repeated freeze-thawing at -80 °C for three times (RE, -14.56%~8.22%), storaged at -80 °C for 30 days (RE, -12.08%~7.56%), and placing them in the sample trays for 24 h after treatment (RE, -7.50%~2.84%) are all able to maintain good stability.

3.2 Pharmacokinetic characteristics of seven ingredients from JHT in normal rats

The concentrations of seven ingredients in the plasma of rats administered four doses (0.7, 1.4, 2.8, 5.6 g/kg JHT) were determined using the established and validated quantitative analytical methods. The mean concentration-time curves are shown in Fig. 2, and pharmacokinetic parameters were calculated using a non-atrial chamber model (Table 1), including $AUC_{0-t}$, $AUC_{0-\infty}$, $C_{max}$, $T_{max}$, $t_{1/2z}$ and $CL_{z/F}$. Meanwhile, a power function model was used to perform a multi-dose linear analysis with the $AUC_{0-t}$ and $C_{max}$, and the results are depicted in Table 2.

In normal rats, all the ingredients reached the first peak within 1 h after administration, which showed a bimodal phenomenon or fluctuating blood concentration in the time curves of each ingredient after oral administration of JHT at the doses of 2.8 and 5.6 g/kg. The changes of blood concentration of allocryptopine and palmatine hydrochloride were only detectable in plasma of the rats that had been given oral doses of 5.6 g/kg, and 2.8 g/kg, respectively; and due to the low concentration, we only obtained the pharmacokinetic parameters of allocryptopine after oral administration of 5.6 g/kg JHT. The $t_{1/2z}$ of each ingredient was less than 4.5 h, which indicated that they were rapidly eliminated in rats. The high exposure ($AUC_{0-t}$) and $C_{max}$ of tetrahydropalmatine, corydaline, costunolide, and rhamnosylvitexin revealed good absorption. The results of linear analysis showed that the $AUC_{0-t}$ and $C_{max}$ of tetrahydropalmatine, corydaline, and costunolide were in a good linear relationship ($r > 0.9$), which indicated that the transient behavior of them in the rats was linear. Whereas, the $AUC_{0-t}$ and $C_{max}$ of rhamnosylvitexin showed dose dependence only when oral administration of 0.7 ~ 2.8 g/kg JHT. The $AUC_{0-t}$ and $C_{max}$ of dehydrocorydaline were relatively low. Meanwhile, similar to rhamnosylvitexin, it also exhibited linear pharmacokinetic behavior after oral administration of 0.7 ~ 2.8 g/kg JHT.

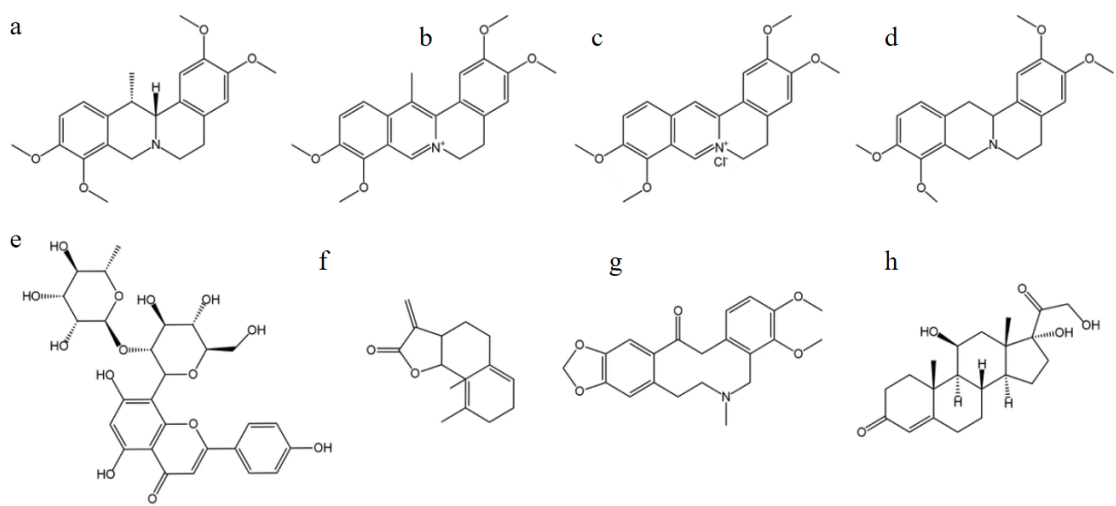

Fig. 1 The chemical structures of seven ingredients in JHT (a. Corydaline, b. Dehydrocorydaline, c. Palmatine hydrochloride, d. Tetrahydropalmatine, e. Rhamnosylvitexin, f. Costunolide, g.

Allocryptopine) and internal standard (h. Hydrocortisone).

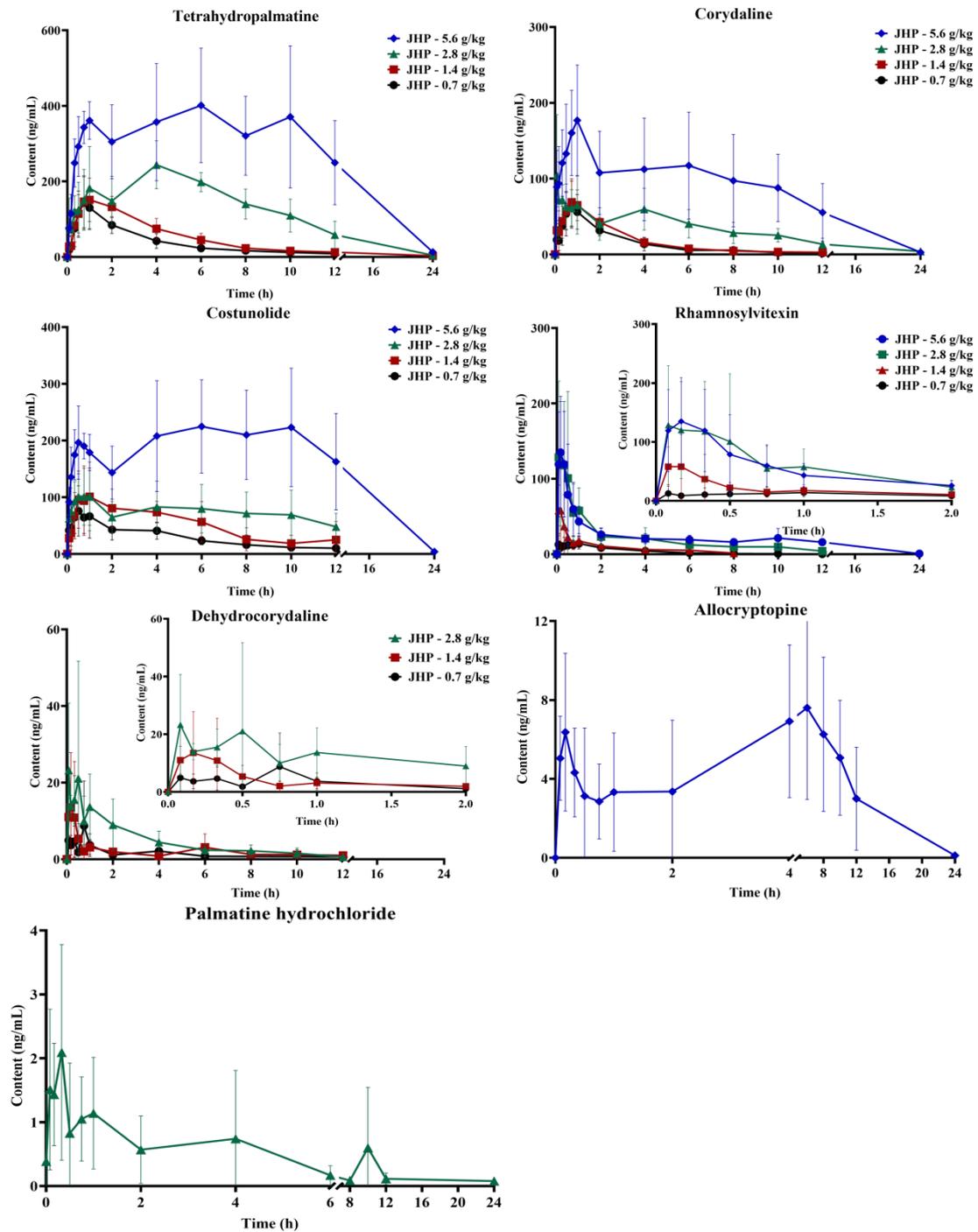

Fig. 2 Concentration-time curves in plasma of seven ingredients in normal rats after oral administration of 0.7, 1.4, 2.8 and 5.6 g/kg JHT ($n$ = 6-8).

Table 1 Pharmacokinetic parameters of seven ingredients in normal rats after oral administration of different doses of JHT ($n$ = 6-8).

| Ingredients | Dose (g/kg) | AUC$_{0-t}$ (μg·h/L) | AUC$_{0-\infty}$ (μg·h/L) | C$_{max}$ (ng/mL) | T$_{max}$ (h) | t$_{1/2z}$ (h) | CL$_{z/F}$ (L/h/kg) |
|---|---|---|---|---|---|---|---|
| Tetrahydropalmatine | 5.6 | 5706.29±1773.89 | 5768.58±1769.23 | 466±147.35 | 6.39±3.12 | 3.06±0.7 | 0.82±0.28 |
|  | 2.8 | 2248.03±399.58 | 2270.46±398.67 | 250.33±58.79 | 4.67±1.03 | 3.1±0.76 | 0.98±0.21 |
|  | 1.4 | 728.9±163.01 | 772.49±161.52 | 165.2±47.53 | 1.44±1.11 | 3.4±0.85 | 1.45±0.28 |
|  | 0.7 | 482.67±105.5 | 528.08±117.51 | 150.74±58.36 | 0.96±0.47 | 3.83±0.7 | 1.09±0.36 |
| Corydaline | 5.6 | 1570.96±833.77 | 1608.54±834 | 188.47±61.22 | 1.45±2.03 | 2.99±0.31 | 4.33±2.18 |
|  | 2.8 | 503.48±107.21 | 557.49±129.94 | 126.1±44.24 | 0.79±1.46 | 3.27±0.94 | 5.14±1.22 |
|  | 1.4 | 212.55±47.57 | 222.44±50.64 | 71.49±25.68 | 0.88±0.55 | 3.22±1 | 6.43±1.42 |
|  | 0.7 | 177.2±23.04 | 188.98±27.13 | 70.4±26.5 | 0.96±0.53 | 4.12±1.22 | 3.67±0.6 |
| Costunolide | 5.6 | 3269.9±1165.39 | 3357.49±1084.64 | 284.17±74.58 | 4.17±3.59 | 2.44±1.14 | 0.25±0.08 |
|  | 2.8 | 888.84±211.93 | 967.78±240.62 | 138.5±33.53 | 3.6±3.88 | 2.91±0.62 | 0.42±0.09 |
|  | 1.4 | 596.26±185.02 | 640.56±220.99 | 122.08±38.87 | 1.79±2.14 | 2.4±0.75 | 0.34±0.1 |
|  | 0.7 | 342.86±36.78 | 383.71±35.8 | 98.12±32.99 | 1.07±1.47 | 3.41±1.39 | 0.26±0.03 |
| Rhamnosylvitexin | 5.6 | 407.02±106.14 | 418.18±96.34 | 142.29±68.16 | 0.18±0.07 | 2.89±1.43 | 25.61±6.16 |
|  | 2.8 | 257.35±69.37 | 272.18±75.86 | 174.83±105.84 | 0.86±1.42 | 2.64±1.74 | 20.48±7.32 |
|  | 1.4 | 76.57±9.01 | 78.68±9.88 | 77.6±53.52 | 0.31±0.2 | 1.36±0.47 | 32.88±4.42 |
|  | 0.7 | 43.17±8.27 | 44.28±8.81 | 21.11±10.4 | 0.73±0.7 | 1.69±0.59 | 30.04±6.68 |
| Dehydrocorydaline | 2.8 | 53.99±17.59 | 57.09±18.1 | 34.95±27.44 | 0.2±0.16 | 2.6±1.22 | 69.09±21.07 |
|  | 1.4 | 24.43±12.48 | 25.32±13.15 | 22.48±13.22 | 0.24±0.15 | 2.56±0.87 | 86.78±36.51 |
|  | 0.7 | 10.78±4.59 | 12.23±4.5 | 9.54±9.04 | 1.17±1.44 | 2.1±0.73 | 84.82±37.45 |
| Allocryptopine | 5.6 | 59.46±31.83 | 69.6±38.62 | 10.49±2.22 | 4.39±3.59 | 3.55±1.32 | 32.13±17.73 |

Table 2 Linear equation and confidence interval between $AUC_{0-t}$, $C_{max}$ and doses of seven ingredients in normal rats after oral administration of JHT.

| Dose range (g/kg) | Ingredients | Parameters | Linear equation | 90% Confidence interval | Judgment interva |
|---|---|---|---|---|---|
| 0.7~5.6 | Tetrahydropalmatine | $AUC_{(0-t)}$ | y = 1435.1x - 571.19 (r=0.9866) | 1.070~1.373 | 0.893~1.107 |
| | | $C_{max}$ | y = 82.276x + 83.359 (r=0.9837) | 0.418~0.766 | 0.828~1.172 |
| | Corydaline | $AUC_{(0-t)}$ | y = 390.77x - 146.95 (r=0.9679) | 0.848~1.264 | 0.893~1.107 |
| | | $C_{max}$ | y = 33.097x + 47.093 (r=0.9769) | 0.308~0.734 | 0.828~1.172 |
| | Costunolide | $AUC_{(0-t)}$ | y = 794.53x - 271.55 (r=0.943) | 0.825~1.207 | 0.893~1.107 |
| | | $C_{max}$ | y = 48.88x + 61.734 (r=0.9473) | 0.342~0.718 | 0.828~1.172 |
| | Rhamnosylvitexin | $AUC_{(0-t)}$ | y = 102.15x - 3.523 (r=0.9536) | 0.999~1.284 | 0.893~1.107 |
| | | $C_{max}$ | y = 29.28x + 44.667 (r=0.5116) | 0.560~1.270 | 0.828~1.172 |
| 0.7~2.8 | Rhamnosylvitexin | $AUC_{(0-t)}$ | y = 146.18x - 52.475 (r=0.9632) | 1.037~1.504 | 0.839~1.161 |
| | | $C_{max}$ | y = 94.191x - 27.5 (r=0.9985) | 0.829~1.975 | 0.743~1.258 |
| | Dehydrocorydaline | $AUC_{(0-t)}$ | y = 27.941x - 3.6574 (r=0.9979) | 0.789~1.540 | 0.839~1.161 |
| | | $C_{max}$ | y = 15.095x + 3.3033 (r=0.9601) | 0.217~1.709 | 0.743~1.258 |

3.3 Pharmacokinetic difference of seven ingredients in normal, CSG and FMT rats

3.3.1 Evaluation of gastric pathological state and intestinal microbial changes

The HE staining results of the rats' gastric tissues are shown in Fig. 3 A and B. Compared with normal rats, the gastric tissues from the CSG model rats showed pathological phenomena such as congestion, histiocyte detachment, inflammatory cell infiltration, etc., and a small amount of hemorrhage was observed in a few gastric tissues; the degree of pathological damage in the model group was significantly higher compared with that in the normal group ($P < 0.05$), indicating that a CSG rat model was successfully established. In the FMT group, some gastric tissues showed congestion, and a small amount of inflammatory cell infiltration was observed in the mucosa of individual rats, but the degree of pathological damage was not significantly different with the normal group.

The intestinal microbial changes are presented in Fig. 3 C, D, and E. Microbial dysbiosis index (MDI) is used to estimate the degree of microbial ecological disturbance, and the greater the value, the greater the degree of microbial disorder. MDI of model and FMT groups were significantly higher than that of normal group ($P < 0.05, 0.001$) (Fig. 3 C). Meanwhile, multivariate statistical results showed that the three groups were completely distinguished (Fig. 2 D). Community heatmap analysis on genus level (Fig. 3 E) demonstrated that the intestinal microbial composition in the FMT group was between the normal and model group.

In short, CSG rats had obvious gastric pathological changes and intestinal flora disorder, while FMT rats only appeared distinctly intestinal flora disorder, indicating the success of CSG modeling and FMT experiment.

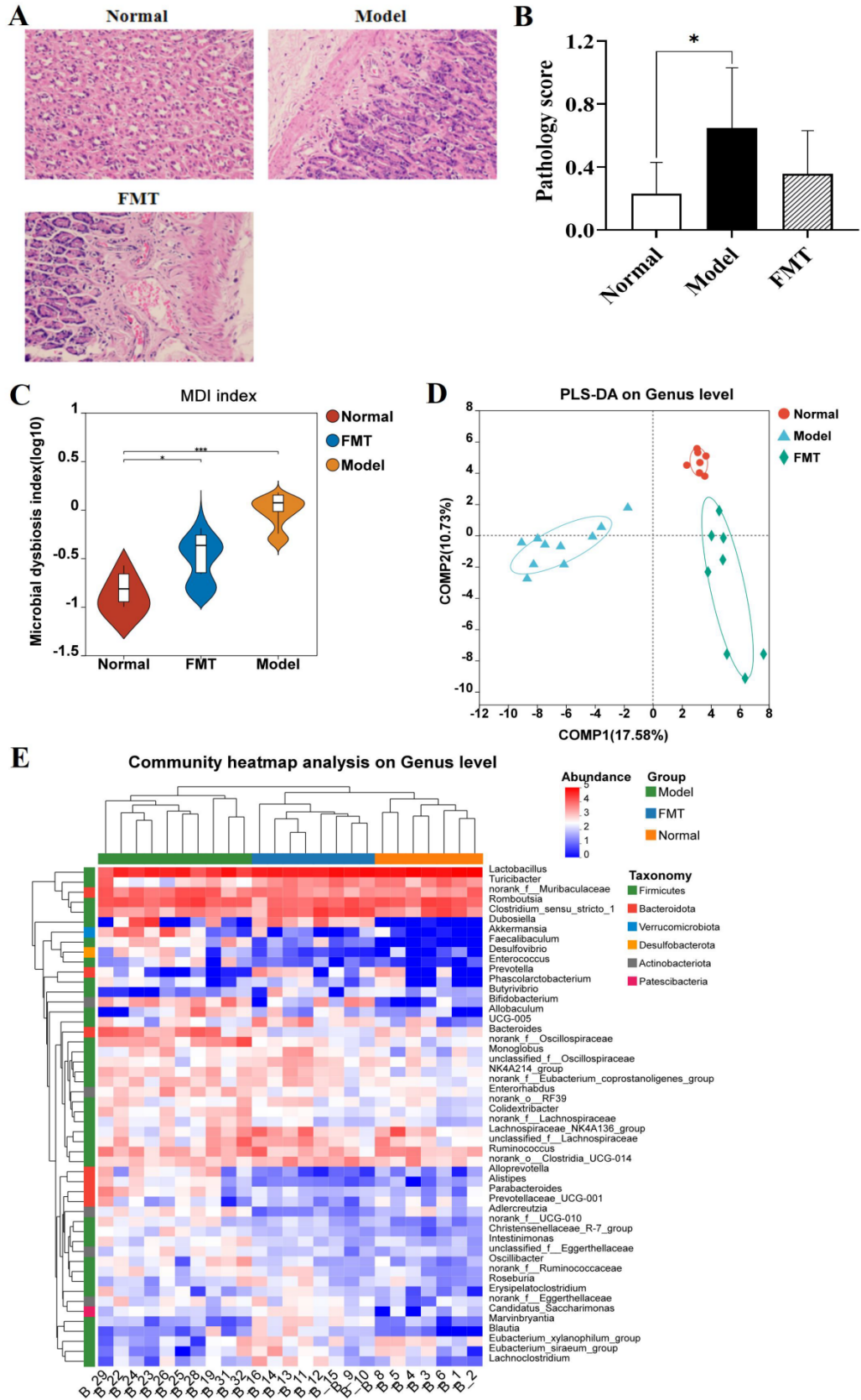

Fig. 3 Gastric histopathologic changes and gut microbiota results in normal, CSG, and FMT rats.

(A) The gistopathological changes in the stomach of normal, CSG, and FMT rats ($n = 8$ for each group). (B) The gastric histopathological damage scores in normal, CSG, and FMT rats ($n = 8$ for each group). (C) The microbial dysbiosis index (MDI) in normal, CSG, and FMT rats ($n = 8$-$10$ for each group). (D) The partial least squares discriminant analysis (PLS-DA) in normal, CSG, and FMT rats ($n = 8$-$10$ for each group). (E) The community heatmap analysis on genus level in normal, CSG, and FMT rats ($n = 8$-$10$ for each group).

3.3.2 Pharmacokinetic comparison of JHT in rats from the three different group.

The concentration-time profiles of the seven ingredients in plasma of normal, CSG, and FMT rats are presented in Fig. 4, and their pharmacokinetic parameters are shown in Table 3. Most of concentrations of dehydrocorydaline, palmatine hydrochloride, and allocryptopine were below their LLOQ, so their pharmacokinetic parameters were not calculated. Comparing normal with CSG and FMT rats, it was found that both disease state and gut microbiota changes had some effects on the absorption and metabolism of JHT, in which the $t_{1/2z}$ of tetrahydropalmatine was significantly decreased in CSG and FMT rats ($P < 0.01$), and the $AUC_{0-t}$ and $AUC_{0-\infty}$ of tetrahydropalmatine was slightly decreased in CSG rats, which was probably due to the shortening of the half-life and the elevation of the clearance rate, leading to its excessive clearance in *vivo*. There were no significant changes in its $AUC_{0-t}$, $AUC_{0-\infty}$, and $C_{max}$. Costunolide had the higher exposure ($C_{max}$, $AUC_{0-t}$, and $AUC_{0-\infty}$) and lower $t_{1/2z}$ ($P < 0.05$, $0.01$) in CSG rats than normal rats. The $t_{1/2z}$ of rhamnosylvitexin was likewise significantly lower ($P < 0.05$). Compared with normal rats, the $AUC_{0-t}$ and $AUC_{0-\infty}$ of FMT rats was significantly high ($P < 0.05$) for corydaline; and the $C_{max}$ was significantly high ($P < 0.01$) and AUC present an increasing trend for rhamnosylvitexin.

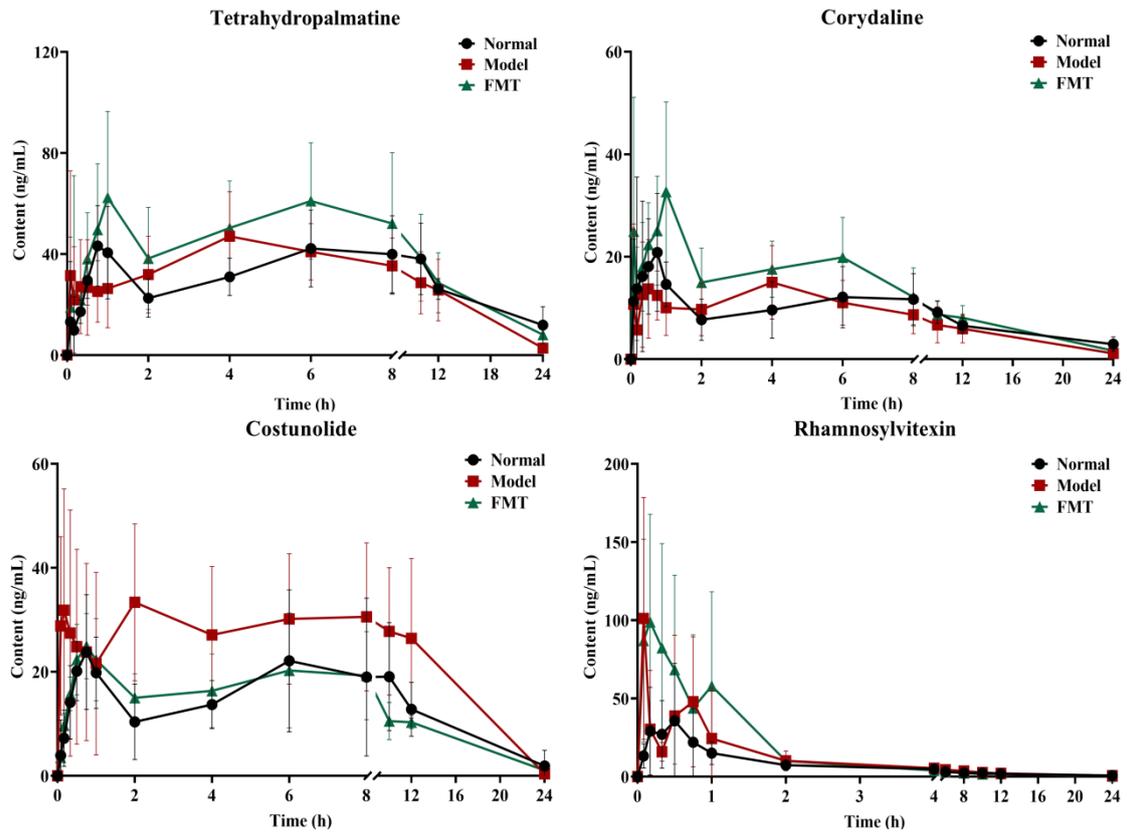

Fig. 4 Concentration-time curves in plasma of four ingredients in normal, CSG, and FMT rats after oral administration of JHT ($n$ = 6-8 for each group).

Table 3 Pharmacokinetic parameters of four ingredients in the plasma of normal, CSG and FMT rats after oral administration of JHT ($n$ = 6-8).

| Ingredients | Group | $AUC_{0-t}$ (µg·h/L) | $AUC_{0-\infty}$ (µg·h/L) | $C_{max}$ (ng/mL) | $T_{max}$ (h) | $t_{1/2z}$ (h) | $CL_{z/F}$ (L/h/kg) |
|---|---|---|---|---|---|---|---|
| Tetrahydropalmatine | Normal | 687.43±200.24 | 869.74±320.33 | 60.85±13.51 | 2.81±3.33 | 8.9±4.1 | 2.8±1.06 |
| | Model | 583.78±150.64 | 601.03±155.69 | 61.06±24.45 | 3.92±3.45 | 4.12±0.73** | 3.77±0.76 |
| | FMT | 779.31±221.52 | 874.42±258.28 | 90.63±31.14 | 2.52±2.65 | 6.62±3.15** | 2.73±1.07 |
| Corydaline | Normal | 183.83±51.74 | 203.8±44.87 | 27.83±18.1 | 2.14±3.07 | 7.06±2.89 | 13.96±3.01 |
| | Model | 159.79±53.36 | 169.02±53.21 | 20.59±10.4 | 1.88±2.45 | 5.62±1.3 | 17.8±6.13 |
| | FMT | 249.62±61.65* | 264.91±60.92* | 39.21±13.86 | 0.64±0.45 | 5.79±1.94 | 10.96±3.41 |
| Costunolide | Normal | 272.46±102.9 | 280.07±100.6 | 27.9±10.64 | 3.04±3.42 | 4.17±2 | 1.53±0.48 |
| | Model | 486.49±178.88* | 487.8±178.72* | 51.38±15.14** | 4.7±4.42 | 2.3±0.26** | 0.92±0.39* |
| | FMT | 242.66±67.3 | 244.92±66.94 | 29.6±6.57 | 3.22±3.47 | 3.25±0.63 | 1.71±0.53 |
| Rhamnosylvitexin | Normal | 84.91±34.52 | 87.29±35.13 | 42.11±37.2 | 0.27±0.14 | 5.46±1.12 | 131.69±44.15 |
| | Model | 114.16±59.74 | 118.11±62.14 | 88.08±67.47 | 0.38±0.35 | 3.78±1.18* | 103.29±39.71 |
| | FMT | 143.13±82.71 | 145.43±83.28 | 133.94±31.85** | 0.3±0.33 | 4.66±1.9 | 89.41±44.77 |

Note: Compared with the normal, *$P$ < 0.05, and **$P$ < 0.01.

3.4 Exploration on the relationship between the *in vivo* exposure of four ingredients and intestinal microbiota

Based on the above results, we characterized the correlation between in *vivo* exposure of four ingredients and 43 disordered gut microbiota in the normal and FMT rats (Fig. S2) by spearman correlation analysis (Fig. 5). The results showed a significant positive correlation ($P < 0.05$ or $P < 0.01$) between corydaline and 17 genera, including g__norank__f__UCG-010, g__Anaerofilum, g__ Anaerostipes, g__Paludicola, g__unclassified__f__Erysipelatoclostridiaceae, g__Frisingicoccus, g__Blautia, g__Anaerovorax, g__norank_f__Butyricicoccaceae, g__Anaerovorax, g__ Christensenellaceae_R-7_group, g__Lachnospiraceae_UCG-001, g__NK4A214_group, g__unclassified __f__Oscillospiraceae, g__Eubacterium_fissicatena_group, g__norank__f__Eubacterium_coprostanoligenes_group, g__unclassified__o_Oscillospirales, g__norank__f__Christensenellaceae. Rhamnosylvitexin was positively correlated with g__DNF00809 and g__Faecalibaculum ($P < 0.05$), and negatively correlated with g__Bacteroides ($P < 0.01$). The result deduced that the gut microbiota changes alter the pharmacokinetic behavior of JHT, especially corydaline and rhamnosylvitexin, which is consistent with the above pharmacokinetic results.

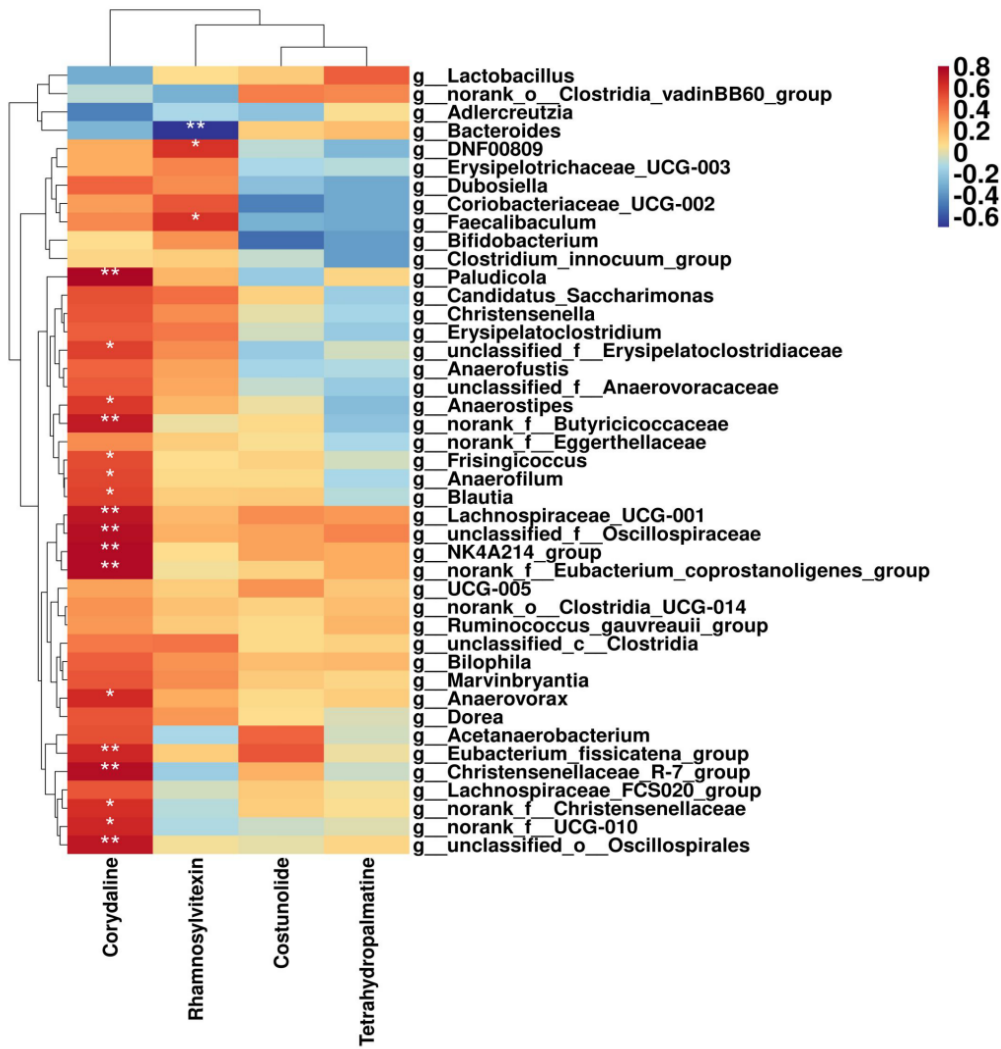

Fig. 5 Heatmap analysis of spearman correlation between four ingredients and 43 disordered gut microbiota in the normal and FMT rats. Red represents positive correlation, and blue indicates negative correlation. *$P < 0.05$, and **$P < 0.01$.

4. Discussion

Alkaloids, as the main active components of yanhusuo, possess significant anti-inflammatory, anti-gastrointestinal ulcer, and analgesic effects, and they can inhibit gastric acid secretion, pepsin activity, and gastric mucosal damage, as well as suppress inflammatory and oxidative stress responses [Xu et al, 2024; Chen, 2021; Gao et al, 2023; Sun and Li, 2024]. Our results indicated that all alkaloid components were rapidly absorbed, which may be attributed to their high solubility, leading to high dissolution rate [Wu et al, 2021]. Among them, tetrahydropalmatine and corydaline had high exposure levels in plasma. After entering the body, only a portion of tetrahydropalmatine is rapidly metabolized, while the majority exists in the plasma in its prototype [Wang et al, 2017], its predicted oral bioavailability is as high as 74%

(https://old.tcmsp-e.com/tcmsp.php). The excretion of corydaline in urine, feces, and bile show that only 18.5% of the administered dose is excreted in its prototype, suggesting that most is metabolized and transformed in the body [Wang et al, 2013]. Therefore, despite the slightly higher content of corydaline than tetrahydropalmatine in JHT, it still exhibited lower plasma exposure than tetrahydropalmatine. On the other hand, quaternary ammonium alkaloids such as dehydrocorydaline and palmatine hydrochloride tend to decompose during the vinegar-processing procedure, resulting in a decrease in their content [Wu et al, 2021]. The strong polarity of quaternary ammonium alkaloids also led to poor absorption of the alkaloids in *vivo* [Lin et al, 2008], explaining the low plasma concentrations of dehydrocorydaline and palmatine hydrochloride in rats. Additionally, it has been reported that no prototype or metabolites were detected in rat plasma after oral administration of allocryptopine duo to rapid absorption and excretion and low bioavailability [Huang et al, 2018]. Moreover, the plasma concentration-time profiles of these alkaloids exhibited a bimodal phenomenon, consistent with the literature [Cui et al, 2018; Tao et al, 2019], which is related to distribution, reabsorption, and enterohepatic circulation [Yang et al, 2013].

Costunolide is a sesquiterpene lactones constituent extracted from chuanmuxiang, and its terpene skeleton and lactone moiety are highly lipophilic, implying good transmembrane permeability and hydrolysis, which is more likely to be translocated out of the cell or metabolized, resulting in a low oral bioavailability [Yu et al, 2021]. However, in multiple experiments involving oral administration of traditional Chinese medicine formulations or muxiang extracts, costunolide showed good bioavailability when oral administration of costus extract, and its $AUC_{0-t}$ and $C_{max}$ were significantly higher than those achieved with the administration of the single compound [Dong et al, 2018; Zhang et al, 2014; Zhang et al, 2015], which may be attributed to the interactions among various complex ingredients, altering their pharmacokinetic profiles. In this study, costunolide exhibited the second-highest $AUC_{0-t}$ and $C_{max}$, following tetrahydropalmatine. Additionally, the plasma concentration-time curves of costunolide showed a bimodal phenomenon, which is speculated to be related to enterohepatic circulation [Hattori et al, 1986]. Studies showed that the concentration of costunolide in colonic tissue increased with time, suggesting that it was absorbed in the intestine [Zhang et al, 2024]. However, some scholars argued that the bimodal phenomenon disappears after intravenous injection of costunolide, thus questioning the reliability of the enterohepatic circulation explanation; instead, they proposed that periodic gastric emptying or discontinuous absorption was a more reasonable explanation for the multi-peak phenomenon [Yu et al, 2021]. However, what we can confirm that a bimodal phenomenon often occurs after oral administration of costunolide, and its specific mechanism needs more supporting data.

Rhamnosylvitexin is a flavonoid compound derived from honghuabajiaoye. Our results showed that it reached the first peak within 10 minutes, with new absorption peaks appearing at 1 hour or 10 hours, and the $t_{1/2z}$ ranged from 1.36 to 2.89 hours, similar to the pharmacokinetic behavior of rhamnosylvitexin after oral administration of shanzhaye extract in rats [Xu, 2019; Ma et al, 2017].

In this experiment, we compared the pharmacokinetic characteristics of four ingredients in normal, CSG, and FMT rats. Compared with normal rats, CSG rats exhibited increased absorption of costunolide and rhamnosylvitexin, whereas the plasma exposure of corydaline and tetrahydrocolumbamine was slightly reduced. Reportedly, corydaline and tetrahydrocolumbamine had high affinity for lesion tissues, and their overall exposure in the target organs is much higher than that in plasma [Ma et al., 2024]. Drug absorption primarily occurs in the gastrointestinal tract, and the gut microbiota under different pathological conditions can affect drug absorption and metabolism. It has been reported that eight isoquinoline alkaloids, including berberine and tetrahydrocolumbamine, were easily converted into demethylated metabolites by the gut microbiota in *vitro* [He et al, 2017]. In another clinical study, it was found that the AUC of three protoberberine alkaloids in the plasma of Chinese individuals was significantly lower than that in Africans, which was attributed to the richer abundance of Prevotella, Bacteroides, and Megamonas in the Chinese gut and more extensive metabolism [Alolga et al, 2016]. Given the similar core structure of isoquinoline alkaloids and the dominant Bacteroides abundance in CSG rats, it is reasonable to speculate that changes in the gut microbiota under pathological conditions and their high affinity for target organs are the main factors contributing to the reduced plasma AUC and accelerated metabolism of tetrahydropalmatine and tetrahydrocolumbamine. Oral administration of ammonia solution in rats can directly damage the gastric mucosa. In this study, we established a CSG rat model by combining intragastric administration of 30% ethanol with intermittent drinking of sodium deoxycholate and ammonia solution. Histopathological results showed significant damage to the gastric tissue cells in rats. Ammonia can increase the pH of the surrounding microenvironment in the organism [Lee et al, 2004], which may lead to the elevated gastric pH in CSG rats. Studies have shown that the absorption of costunolide and rhamnosylvitexin is affected by the pH of the physiological environment, and gastric acid can cause the degradation of costunolide [Li et al, 2018], rhamnosylvitexin show optimal absorption under near-neutral conditions compared to pH 4.0 or pH 8.0 [Chow et al, 2025]. The elevated gastric pH may enhance the stability of these two compounds in the stomach. As the main binding site for the metabolism of lignan lactone, α-methylene-γ-butyrolactone is highly susceptible to carboxylation and sulfhydryl binding, and this type II metabolism can be carried out under physiological pH conditions [Yu et al, 2021], suggesting that the elevation of intragastric pH may inhibit this metabolism process to some extent. In addition, chronic ethanol exposure can upregulate the expression of P-glycoprotein (P-gp) [Gong et al, 2020]. P-gp is the most important drug efflux pump in the parietal membranes of many various cells, such as intestinal epithelial cells [Elmeliegy et al, 2020]. It has been demonstrated that lactone components have binding activity with P-gp, and rhamnosylvitexin show significant transporter-mediated intestinal efflux [Chow et al, 2025], and more desirable binding sites for drug efflux proteins may accelerate the elimination of both compounds in *vivo*.

The absorption of JHT was altered in rats after FMT, mainly manifesting in the AUC increase of corydaline and rhamnosylvitexin in FMT rats. Disordered intestinal flora in disease state could weaken the metabolic ability of drugs and increase the permeability of intestinal barrier, leading to the increase of drug prototype absorption [Javdan et al, 2020]. During FMT, the microbiota composition, microbial diversity, and the complementarity between donor and recipient microbiota determined the resilience, coexistence, and colonization of FMT [Schmidt et al, 2022]. In the present study, pharmacokinetic characteristics of JHT were different from that of the donor rats (rats with CSG), which likewise denies the "super donor" hypothesis.

The differential bacterial genera between FMT rats and healthy rats were compared to conducted a correlation analysis with i*n vivo* exposure of four ingredients. Corydaline and rhamnosylvitexin had the positive correlation with the relative abundance of most differential genera, with corydaline showing stronger significance. *Bacteroides*, a beneficial bacterial genus, not only metabolize primary bile acids to a variety of secondary bile acids, forming a bi-directional regulation with the gut microbiota to inhibit the overgrowth of bacteria in the intestine, but also express a series of hydrolases to participate in the metabolism of drugs [Li et al, 2021]. For instance, α-L-rhamnosidase and β-rutinosidase from *Bacteroides sp. 45* could transform rutin into quercetin 3-O-glucoside, quercetin, and leucocyanidin; *Bacteroides sp. 54* could hydrolyze quercitrin to hydroxyquercitrin and desmethylquercitrin; *Bacteroides thetaiotaomicron VPI-5482* could express α-L- rhamnosidase and metabolize epimedin C into icariside [Ferreira-Lazarte et al., 2021; Jiang et al., 2014; Wu et al., 2018; Zhao et al., 2022]. In this research, rhamnosylvitexin showed a highly significant negative correlation with the abundance of *Bacteroides* ($P < 0.01$), indicating that the increased absorption of rhamnosylvitexin is closely related to the decreased abundance of *Bacteroides* in FMT rats. Therefore, it is speculated that the decreased *Bacteroides* leads to the decreased hydrolysis of vitexin rhamnoside, thus increasing the absorption of its prototype. However, the absorption of costunolide remained unchanged, consistent with the previous study that it was not metabolized by intestinal flora.

Additionally, *Faecalibacterium* had significantly higher abundance in patients with Crohn's disease and ulcerative colitis, making it a key microbial genus associated with these diseases; similarly, *Dorea* is significantly elevated in children with non-alcoholic steatohepatitis and patients with irritable bowel syndrome [Juárez-Fernández et al, 2023; Brunkwall et al, 2021]. *Lactobacillus*, one of the common probiotics, is often employed in the prevention and treatment of inflammatory and infectious intestinal diseases [El-Baz et al, 2020] and assist in the recovery of the host's disturbed gut microbiota under intestinal diseases, demonstrating positive therapeutic effects in clinical studies of patients with ulcerative colitis [Huang et al, 2022]. Moreover, *Lactobacillus* can alter levels of short-chain fatty acids and other metabolites to promote the proliferation of intestinal epithelial cells and enhance the barrier capacity of the intestinal mucosa [Wang et al. 2020]. In this study, compared with normal rats, FMT rats have the decreased *Lactobacillus* and increased *Faecalibacterium* and *Dorea*, suggesting that FMT rats may experience intestinal barrier damage and lead to the increase of corydaline absorption.

## 5. Conclusion

In summary, this study established a stable and accurate UPLC-TQ-MS/MS quantitative analytical method for seven ingredients of JHT and proved JHT exhibited the linear pharmacokinetic characteristics within the appropriate dose ranges in normal rats. Additionally, the CSG and FMT models were successfully developed to compare the pharmacokinetic behavior of JHT in normal, CSG, and FMT rats, with preliminarily exploration of the intestinal microbiota affecting its absorption and metabolism. The results showed that the CSG pathological state affected the absorption and metabolism of costunolide and rhamnosylvitexin; FMT increased that of corydaline and rhamnosylvitexin, closely related to 20 differential bacterial genera. These suggest that both gut microbiota structure and gastric pathological changes can affect the pharmacokinetic of JHT. We investigate the pharmacokinetic characteristics of JHT at four doses in rats and explored the impact of pathological conditions and simple intestinal microbial disorder on the absorption of JHT for the first time.

**Declaration of competing interest**

The authors confirm that there are no conflicts of interest.

**Data availability**

Data will be made available on request.